\documentclass[aps,prx,twocolumn,english,showpacs,letterpaper,superscriptaddress]{revtex4-1}

\usepackage[colorlinks=true,citecolor=blue,linkcolor=blue]{hyperref}

\usepackage[utf8]{inputenc}
\usepackage{amsmath}
\usepackage{amssymb}
\usepackage{amstext}
\usepackage{amsopn}
\usepackage{amsfonts}
\usepackage{amsxtra}
\usepackage{fixltx2e}
\usepackage[english]{babel}
\usepackage{graphicx}
\usepackage{float}
\usepackage{bm}
\usepackage{multirow}
\usepackage{dcolumn}
\usepackage{color}
\usepackage{hyperref}
\usepackage{todonotes}
\usepackage{verbatim}
\usepackage{soul}
\usepackage{xcolor}

\newcommand{\ignore}[1]{}
\newcommand{\mater}{ZrTe$_5$~}

\newcommand{\COMMENTED}[1]{}

\def\mater{ZrTe$_5$}

\date{\today}

\begin{document}
\author{D. Nevola$^{\star}$} 
%\footnote{TEST}
\affiliation{Condensed Matter Physics and Materials Science Division, Brookhaven National Laboratory, Upton, New York 11973, USA}
\author{N. Aryal$^{\star}$}\email[]{naryal@bnl.gov}
\affiliation{Condensed Matter Physics and Materials Science Division, Brookhaven National Laboratory, Upton, New York 11973, USA}
\author{G.D. Gu}
\affiliation{Condensed Matter Physics and Materials Science Division, Brookhaven National Laboratory, Upton, New York 11973, USA}
\author{P.D. Johnson}
\affiliation{Condensed Matter Physics and Materials Science Division, Brookhaven National Laboratory, Upton, New York 11973, USA}
\author{W.-G. Yin}\email[]{wyin@bnl.gov}
\affiliation{Condensed Matter Physics and Materials Science Division, Brookhaven National Laboratory, Upton, New York 11973, USA}
\author{Q. Li}\email[]{liqiang@bnl.gov}
\affiliation{Condensed Matter Physics and Materials Science Division, Brookhaven National Laboratory, Upton, New York 11973, USA} \affiliation{Department of Physics and Astronomy,
Stony Brook University, Stony Brook, New York 11794-3800, USA}

%\title{Phonon-Mediated Band Structure Softness in ZrTe$_5$}
\title{Origin of light-induced metastability in ZrTe$_5$}

\begin{abstract}
We study the non-equilibrium electronic structure of a model Dirac semimetal \mater~ by using time-and-angle resolved photoemission spectroscopy and density functional theory-based electron and phonon calculations. By measuring the electronic dispersion near the $\Gamma$ point at time delays up to 10 picoseconds, we discovered that the band spectral weight does not recover during the measured temporal window,  revealing the existence of light induced metastable state in the electronic structure of this material.
%while most of the spectral weight recover after couple of picoseconds, the spectral weight at the top of the valence band around $\Gamma$ point 
Our calculations find that the photoexcited $A_{1g}$ phonon mode  
%involving zig-zag Te atoms 
lead to a band renormalization that both supports our experimental observations at the zone center and predicts changes to the band structure outside of our experimental window, ultimately showing the evolution from a direct to an indirect gap semimetal; 
such band renormalization dramatically reduces the electron-hole recombination rate giving rise to the metastability in this system.

%\red{which  dramatically reduces the electron-hole recombination rate giving rise to the metastability in this system.} 
\end{abstract}

\maketitle

%\def\thefootnote{\star}\footnotetext{fadsa}
%\makeatletter
%\def\blfootnote{\gdef\@thefnmark{}\@footnotetext}
%\makeatother

%\thefootnote
%\footnote{dasa}

\textit{Introduction.--}
Metastable states are crucial components for quantum computers including those built on trapped ion processors and semiconductor spin qubits~\cite{metastable_Kang_PRX2023,omg_metastable_Allcock_APL2021}. On the other hand, the fragile metastable states are also a primary source of decoherence producing errors in quantum computers. In most of quantum information systems, metastable states are manipulated through interaction with environment – mostly by light (or electromagnetic waves at a specific frequency)~\cite{Erasure_Wu_NatComm2022,dualqubit_Yang_NatPhys2022}. In recent years, applications of topologically protected states have been considered for error-tolerant quantum computers~\cite{chiralqubit_Zhang_PNAS2018,kharzeev2019chiral}. A grand challenge underlying the field of topology-enabled quantum logic and information science is how to establish control principles of topological quantum states that are driven by light. Thus, it is a prerequisite to understand the light-induced metastability at controllable time scale. Topological Dirac semimetal ZrTe$_5$ has recently emerged as a candidate for topological quantum system because of its light-induced Dirac-Weyl state having a respectable long coherence time (100 ps) without application of static electric or magnetic fields~\cite{ultrafast_switch_Vaswani_PRX2020}.

ZrTe$_5$ has been the subject of intense research in the topological condensed matter community for almost a decade due to the discovery of novel properties such as the chiral magnetic effect~\cite{CME_QiangLi_Nature2016}, 3D quantum~\cite{3DQHE_Tang_Nature2019} and the anomalous hall effect ~\cite{AHE_Liang_NaturePhys2018,AHE_Liu_PRB2021,AHE_Lozano_PRB2022} etc.
Monolayer ZrTe$_5$ is a quantum spin hall insulator~\cite{QSHZrTe5_PRX_Weng_Dai_2014,Edgestates_Wu_PRX2016}, whereas the bulk system is a small gap model Dirac system. 
%Although it is still debated whether the low temperature phase of this system is a strong (STI) or a weak topological insulator (WTI)~\cite{STI_manzoni_PRL2016,ZrTeARPES_WeakTI_XiongPRB2017}, 
%It has been shown by many groups that
External perturbations like strain~\cite{strain_Mutch_sciadv2019}, temperature~\cite{TempDrivenTopology_Xu_PRL2018}, field, phonons~\cite{ultrafast_switch_Vaswani_PRX2020,photocurrent_zrte5_Luo_naturemat_2021} etc. have been shown to induce topological phase transition from strong to weak topological insulator thereby closing and opening of a Dirac mass gap.
%In addition to its unique topological properties, a resistivity anomaly has been seen near 50-120K ($\approx$ 75K in our samples)~\cite{GiantResistivityAnomalyZrTe5_Okada_JPSJ_1980}, which has been attributed to the change in the chemical potential from electron-like to hole-like~\cite{LifshitzZrTe5_Zhang_NatureComm2017, ZrTe5Lifshitz_QLi_IOP2017}. However, it remains to be understood if the Dirac nature of the band dispersion and temperature induced topological phase transition has any relation to the resistivity anomaly~\cite{Diracpolarons_Fu_PRL2020}.
The softness of the band structure to various experimental knobs in this material demonstrates the potential for perturbation with ultrafast optical pulses~\cite{konstantinova2020photoinduced, UltrafastHotCarriers_Zhang_PRB2019}. 
Manzoni \textit{et al.} demonstrated ultrafast shifting of the bands on femtosecond timescale, reminiscent of the Lifshitz transition~\cite{UltrafastControl_Manzoni_PRL2015}. Additionally, ultrafast studies of ZrTe$_5$ have shown the importance of electron-phonon coupling and phonon anharmonicity~\cite{HotCarrierDynamics_Li_PRB2020}.

Recently, Konstantinova \textit{et al.} used bulk sensitive MeV ultrafast electron diffraction measurements~\cite{konstantinova2020photoinduced} to study photoinduced atomic dynamics of ZrTe$_5$. They found unexpectedly large relaxation time of some of the Bragg peaks, of the order of $\sim$100 ps, indicating phonon-driven metastability in this system. 
While it was shown by previous Density Functional Theory-based calculations that such phonon driven structural changes  can modify the electronic bands and induce topological phase transition~\cite{aryal2020topological,aryal2022robust}, experimental measurement of the electronic spectra in the metastable phase and theoretical understanding of the driving mechanism for such metastability are missing.
Owing to its relevance in ultrafast switching and topological quantum computers~\cite{UltrafastSymmetrySwitchWeyl_SieNature2019,ultrafastreview_Weber_JAP2021}, it is important to understand the origin and consequence of such metastability in the picosecond time domain.

 In this work, we study the  electronic structure of \mater~ at picosecond time-delays by using time-and-angle resolved photoemission spectroscopy (trARPES).
We complement our experimental studies by DFT-based electron and phonon calculations.
We find that the band spectral weight close to the valence band maxima in the vicinity of the $\Gamma$ point does not recover within 10 ps, suggesting strong electron-phonon coupling in light of the experimental findings of phonon driven metastability~\cite{konstantinova2020photoinduced}.
A previous all-optical ultrafast experiment has identified that one $A_{1g}$ phonon mode, which involves distortions of the Te atoms, couples the most strongly with the excited electrons during their relaxation~\cite{HotCarrierDynamics_Li_PRB2020}.
Our calculations confirm that this is indeed the most likely non-equilibrium electron-hole relaxation mechanism, and we find that this phonon distortion causes electronic band renormalization. More specifically, the conduction band minimum can shift away from the equilibrium position such that the electron-hole recombination rate is severely suppressed.

%It initially provided the first demonstration of the photogalvanic effect, where a spin imbalance at the Fermi level upon the application of a magnetic field led to potential applications into spintronics. An anomoly in the resistivity was later attributed to a Lifshitz transition, whereby the Fermi surface changes from hole type to electron type at the $\Gamma$ point with an increase in temperature. 

\begin{figure}[tb]
\includegraphics[width=0.5\textwidth]{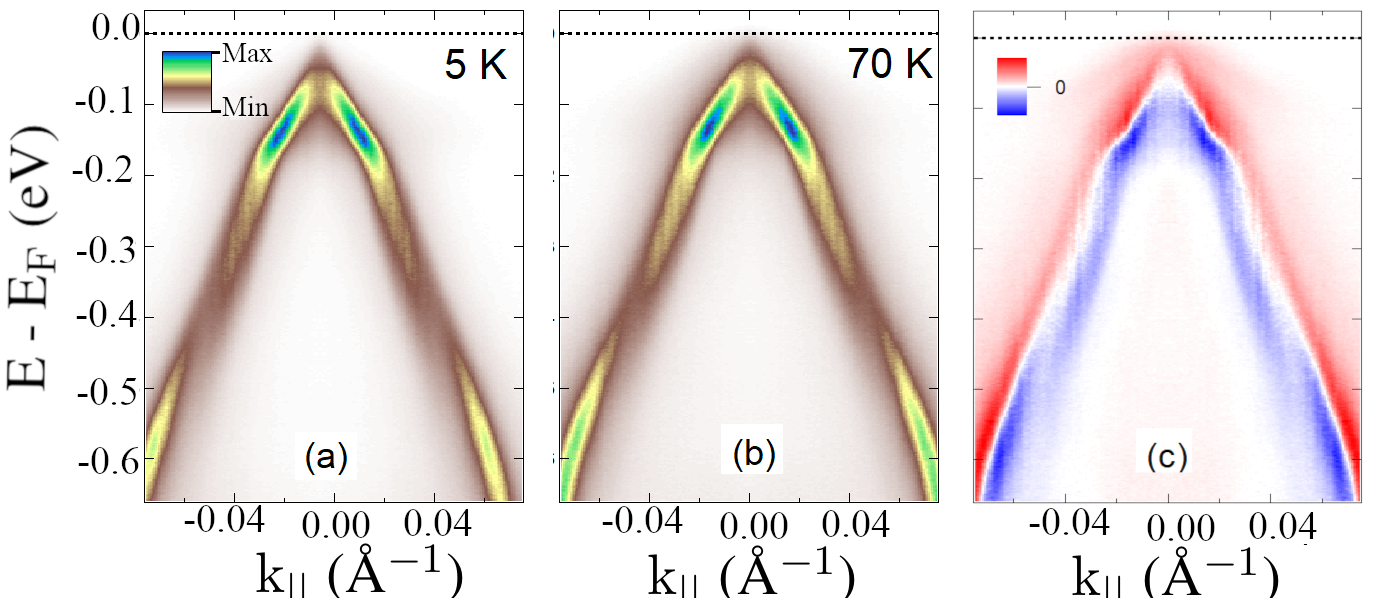}
\caption{High resolution, static ARPES measurements taken at (a) 5~K and 70~K. (c) The difference between (a) and (b), demonstrating the Lifshitz transition. The dashed lines show the location of the Fermi level.}
\label{Fig1}
\end{figure}

\textit{Results.--}We start by reviewing the temperature dependent, equilibrium band structure of \mater. Figs.~\ref{Fig1}(a, b) show the static ARPES results at 5~K and 70~K, respectfully. There are two energy windows of interest that have already been discussed in depth in Refs.~\cite{ARPES_Moreschini_PRB2016,LifshitzZrTe5_Zhang_NatureComm2017,ARPES_Zhu_PRB2022}: below and above $-0.2$~eV. Below this energy, we observe strong intensity variations and two peaks in the momentum distribution curves (see Supplemental Material~\cite{Supplementary}) that have already been attributed to final state effects~\cite{ZrTeARPES_WeakTI_XiongPRB2017}. Above $-0.2$~eV, we see an intensity increase of the energy bands, a peak broadening, and a kink. The peak broadening and intensity in this energy region was previously shown to be sensitive to the photon energy used to probe the electrons. In particular, the band structure probed in this region contains a $k_z$ projection on to the in-plane momenta, where each relative k$_z$ intensity is heavily dependent on the probe photons used. As discussed in Ref.~\cite{ZrTeARPES_WeakTI_XiongPRB2017}, this is likely a combination of matrix element and final state effects, and is an important point that we will return to when discussing our band structure calculations.

\begin{figure*}[htb]
\includegraphics[width=0.95\textwidth]{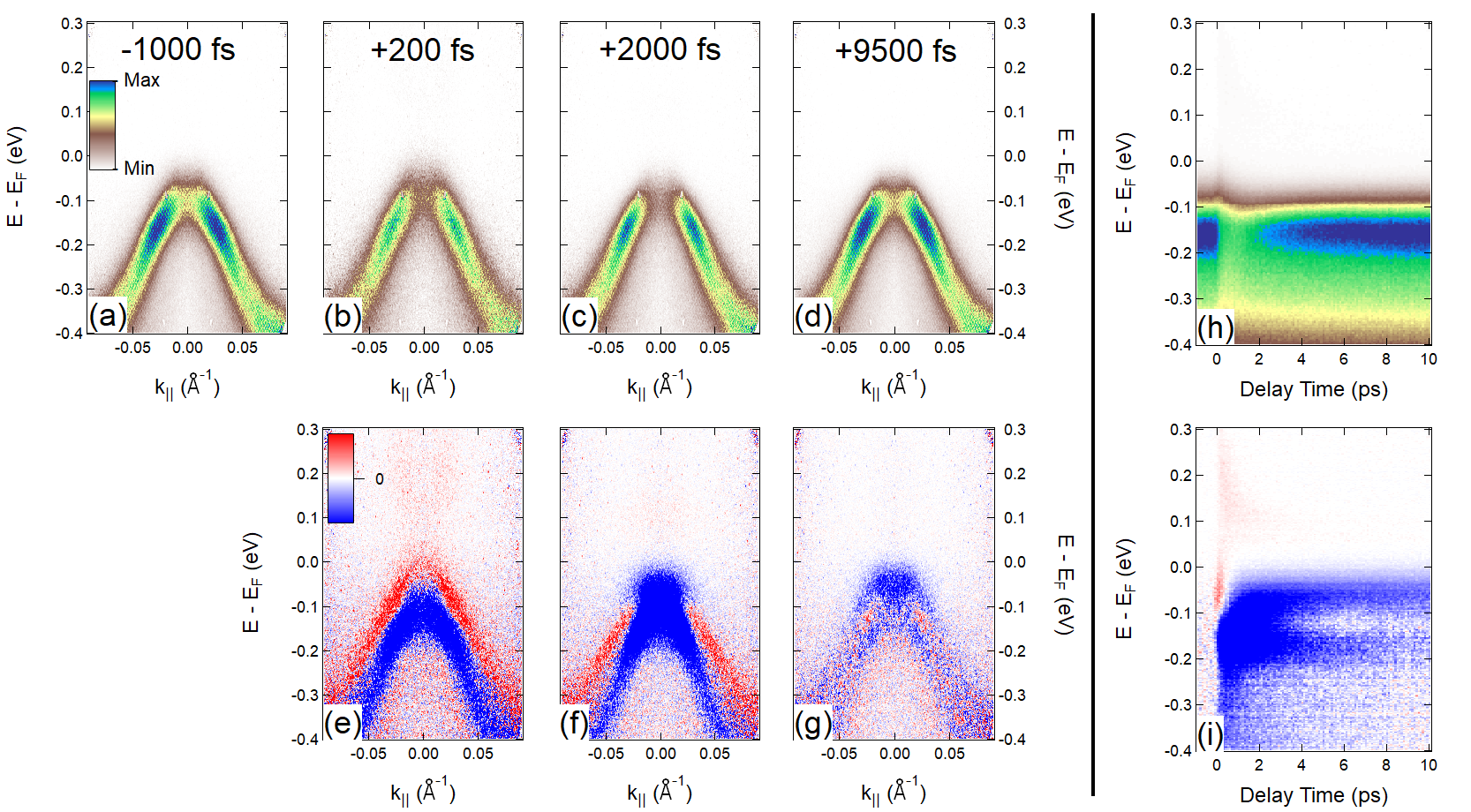}
\caption{Pump-probe ARPES results taken at 5~K and pumped with a fluence of 60~$\mu J/cm^2$. (a-d) Photoemission spectra at the indicated delay times. (e-g) Difference photoemission spectra after subtraction with the spectra at negative time delay (panel (a)). (h) $k$-integrated spectra as a function of delay time. (i) $k$-integrated difference spectra after subtraction at negative delays.}
\label{Fig2}
\end{figure*}

The energy bands are known to undergo a shift with temperature that we highlight in Fig.~\ref{Fig1}. In going from 5~K to 70~K, the bands shift upward by 10~meV. This agrees well with several other studies that have quantified a similar shift of $\approx$ 0.2meV/K~\cite{LifshitzZrTe5_Zhang_NatureComm2017}. The present material undergoes a Lifshitz transition at 70~K, above which, the carriers shift to being electron-like, meaning that the Fermi level crosses the conduction band. 

The band shift is highlighted in Fig.~\ref{Fig1}c, which shows the difference spectrum (70~K - 5~K). Red and blue colors indicate an intensity increase and decrease respectively. Importantly, this shift is uniform across the entire observation window. Given how apparent the shift is in the difference spectrum, we use it as a signature for the Lifshitz transition. The difference spectrum was similarly used as a signature in Ref.\cite{UltrafastControl_Manzoni_PRL2015}, which first used it to report a pump-induced band shift that we also report on here.

\begin{figure}[tb]
\includegraphics[width=0.50\textwidth]{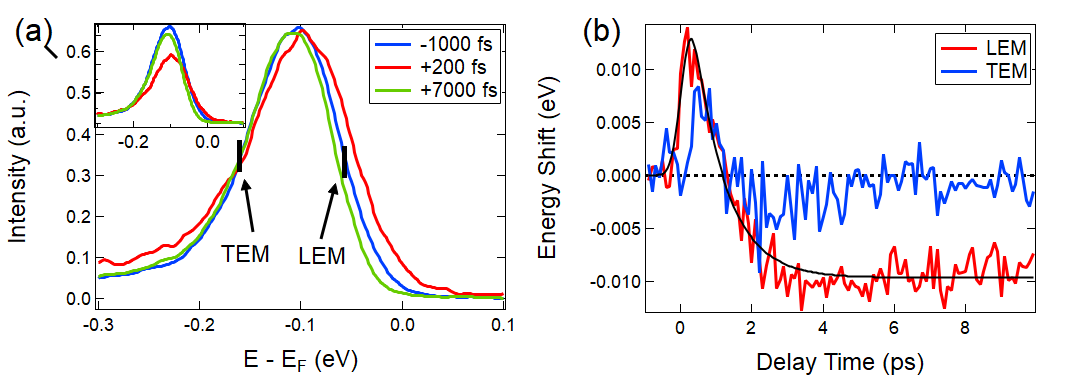}
\caption{Shifting of the valence band. (a) Energy distribution curves taken at the $\Gamma$ point for three of the delay times in Fig.~\ref{Fig2}. The data has been normalized to each other for clarity. The black dashed show the approximate locations of the leading edge midpoint (LEM) and trailing edge midpoint (TEM) as a guide for the reader. (b) Change in the LEM and TEM as a function of delay time. The black line shows the best fit discussed in the main text.}
\label{Fig3}
\end{figure}

Fig.~\ref{Fig2} shows the effects of pumping the system. Fig.~\ref{Fig2}(a-d) shows the spectral intensities at 4 different delay times. Similarly, Fig.~\ref{Fig2}(e-g) shows corresponding difference spectral intensities after subtraction with the spectral intensity at negative time delay. The most obvious change is the intensity decrease at +200~fs, consistent with pumping electrons out of the valence band,
 and subsequent recovery at later delays resulting from quantum scattering processes.
 %\red{Need some citations here}. 
 Similar to the static ARPES data (Fig.~\ref{Fig1}c), we analyze the difference spectra in order to gain a more detailed insight into the pump-induced changes to the band structure. At +200~fs, we observe changes to the valence band that are reminiscent of the data in Fig.~\ref{Fig1}c, indicative of the ultrafast band shift that was first reported in Ref.~\cite{UltrafastControl_Manzoni_PRL2015}. Additionally, we observe a weak population above the Fermi level (E$_\mathrm{F}=0$) that is consistent with the location of the conduction band. 
 
At the later delay time of +2000~fs, we observe an additional hole population at the valence band maxima in addition to the remnant of the ultrafast band shift. This hole population persists up to the maximum observed delay time of 9500~fs, even after the band shift at higher binding energies relaxes back to its equilibrium position.

The time dependence of the band structure changes are summarized in the $k$-integrated data, shown in Figs.~\ref{Fig2}(h, i). In the $k$-integrated difference spectra shown in Fig.~\ref{Fig2}(i), we see an intensity increase at short delays (within the first picosecond) just below the chemical potential that we attribute to the ultrafast band shift. This is followed by persistent intensity decrease that is only present within 200~meV of the Fermi level. At binding energies higher than this, the holes are almost completely recovered within our delay window. Thus, we believe this metastable phenomena is separate from the ultrafast band structure shift that we and other studies observe at short delay times, and it is this feature that we focus on for the remainder of the present study.
 
Figs.~\ref{Fig3}(a,b) show the energy distribution curves (EDC's) at $k_{||}=0$ for three different delay times. We normalize the individual curves as we are interested in the relative positions of the peaks (the non-normalized curves are shown in the inset. The EDC at $+200$~fs is shifted upwards with respect to that at $-1000$~fs, which again shows the ultrafast band shift. However, at the longer delay of $+7000$~fs (green curve), we observe that the trailing edge (the rising edge between $-0.2$~eV and $-0.1$~eV) is nearly unchanged, but the leading edge (the falling edge between -0.1~eV and E$_\mathrm{F}$) is shifted downwards. Thus, the missing spectral weight at the top of the valence band is mostly responsible for the metastable signal previously discussed.

The time dependence of these changes are shown in Fig.~\ref{Fig3}b, where we track the positions of the leading edge midpoint (LEM) and trailing edge midpoint (TEM). The leading edge midpoint shifts upwards at short delays and then downwards, bypassing its original position at negative delays, ultimately settling on a metastable position 10~meV below its initial one. The TEM displays a remarkably different behavior, shifting upwards at short delays and then relaxing back to its initial position after $\sim1.5$~ps. 

\textit{Discussion.--}There are several possibilities for the observed photoemission behavior that we now discuss in detail. The first possibility is that the holes are relaxing to the top of the valence band and are unable to combine with the electrons at the bottom of the conduction band in the measured time window. Another possibility is the change in photoemission matrix elements after pumping or band structure renormalization resulting from the increased phonon population after photoexcitation.

First, we discuss the simple explanation that the holes relax to the top of the valence band and electrons to the bottom of the conduction band. A detailed photoemission study by Xiong \textit{et al.} showed that the valence band at the $\Gamma$ point contains a projection of all the $k_z$ states onto the $k_{||}$ plane, with the relative weight of each $k_z$ projection highly dependent on the photon energy used to probe the system which is indicative of final state effects~\cite{3Dbands_Xiong_PRB2017}. The LEM then contains most of the information at the $\Gamma$ point and the TEM contains most of the information at the the other high symmetry $k_z$ point (see Supplemental Material~\cite{Supplementary}).
%- \textcolor{red}{Niraj, can you include band structure calculations without the kz projections in the supplemental to show this}). 
This would support the explanation of holes relaxing to the top of the valence band at $\Gamma$ being unable to combine with the excited electron population.

However, it has been shown that the conduction band minimum  lies at the $\Gamma$ point, and hence the excited electrons fall within our observation window. As shown in Fig.~\ref{Fig2}(i), the excited electrons fully decay after the first few picoseconds, contradicting this point. It is also possible that electrons are at another local minima outside of our observation window, and are unable to recombine due to an inability to scatter down to the minimum at the $\Gamma$ point. Another possibility we will discuss below is that the band structure is renormalized such that the conduction band minima is no longer at the $\Gamma$ point, which may explain the observed metastable behavior.

The other possibility for the missing spectral weight around $\Gamma$-point is the change in the photoemission matrix elements or a possible change to the final states that can affect the relative intensities of the $k_z$ states. As discussed above, the observed band structure is sensitive to final state effects or matrix elements effects. Therefore, it is possible that the matrix elements at the top of the band change as a result of electronic redistribution and excited phonon populations. Although this is difficult to eliminate this possibility, it is the increased phonon population that would also cause the change to the band structure, as we now discuss.

Finally, we discuss the change to the electronic band structure renormalization as a result of the increased phonon density resulting from photoexcitation. It is well-known that (1) the decay resulting from photoexcitation leads to an increased phonon population and (2) the electronic structure of ZrTe$_5$ is sensitive to temperature (i.e. Lifshitz transition), from which it has been speculated that electron-phonon coupling plays a role.  It is this point that we discuss in detail in the following section. Based on DFT calculations, we argue that the electronic structure is sensitive to the phonon population, which can significantly alter the electron-hole recombination rate. 

\begin{table}[htb] %[!tht]
\begin{center}
%\caption{Displacement vector of A$_g$-25 mode.}
    \caption{ Results from constrained DFT relaxation calculations performed at different values of electronic  temperature. The change in atomic coordinates from the equilibrium structure is expressed in terms of the normal mode coordinates ($Q$ values) corresponding to the 6 $A_{1g}$ Raman phonon modes~\cite{aryal2020topological}. The unit of $Q$ is  \AA$\sqrt{\mathrm{amu}}$.}
\begin{tabular}{|p{28pt}|p{30pt}|p{32pt}|p{32pt}|p{32pt}|p{32pt}|p{32pt}|} \hline
 T$_{e}$(K) & Q$_{A1g-6}$ & Q$_{A1g-22}$ & Q$_{A1g-25}$ & Q$_{A1g-27}$ & Q$_{A1g-31}$ & Q$_{A1g-36}$\\ \hline
2000  & 0.016 	& 0.036 & 0 & 0.32 & 0.013 & 0\\ \hline	
4000 & 0.006	& 0.032	& -0.08	 & 0.84 & 0.23 & 0 \\ \hline	
6000 & -0.1 	& 0.14	& -0.41 & 1.82 & 1.1 & 0 \\ \hline	
\end{tabular}
\label{table:T_elec}
\end{center}
\end{table}

\begin{figure}[htb]
\includegraphics[width=0.47\textwidth]{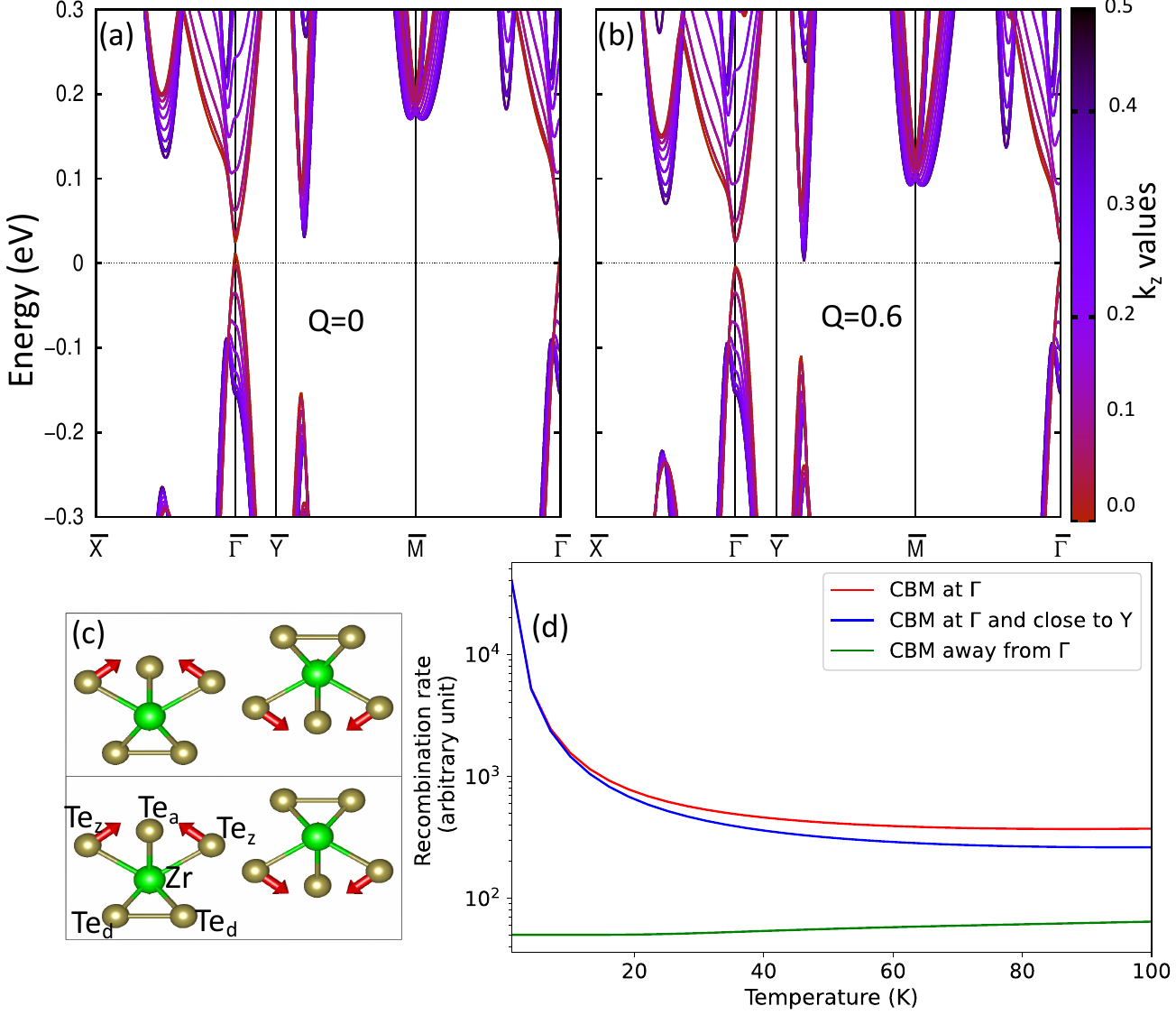}
\caption{(a \& b) Calculated $k_z$ projected band structure of ZrTe$_5$ for equilibrium structure and A$_{1g}$-27 phonon mode distorted structure with $Q$ value of 0.6 (in units of \AA$\sqrt{amu}$), respectively. 
(c)$A_{1g}$-27 phonon mode distortion.
(d) Estimation of electron-hole recombination rate as a function of temperature using Eq.~\ref{eq:recombination_rate} for different positions of conduction band minima.
}
\label{Fig4}
\end{figure}

When the system is pumped by an optical pulse, experiments~\cite{HotCarrierDynamics_Li_PRB2020} indicate that one phonon mode, corresponding to the displacement of Te atoms along the zig-zag direction (called Te$_z$ here), is dominantly excited. 
From our phonon calculations, we identify this phonon mode to be $A_{1g}$-27 phonon mode with frequency of $\sim$4.3 THz.
%From ultrafast electron diffraction experiment as well as \blue{other experiments},  
In addition, there are indications~\cite{konstantinova2020photoinduced} that the system is driven into a metastable state corresponding to increased phonon populations during the ultrafast relaxation of hot carriers.
In order to understand this phenomena, we perform lattice relaxation calculations at different values of electronic temperature T$_e$ by preserving lattice symmetry~\cite{VO2_WallScience2018,aryal2024origin}. In this approach, we create electron (hole) occupations in the equilibrium conduction (valence) bands by Fermi-Dirac distribution at the given T$_{e}$ and the atomic positions are relaxed under this constraint. 
Relaxation of atomic coordinates by preserving lattice symmetry corresponds to the $A_{1g}$-type phonon-mode lattice distortions and there are six such phonon modes in this \mater~\cite{aryal2020topological}.
Since the high-energy photo-excited electrons and holes relax quickly to the bottom of the conduction and valence bands within the first few hundred femtoseconds~\cite{Bovensiepen2012},  this approach, though ad-hoc, can provide insights into how the the lattice responds to the change in electron and hole population over a larger (picosecond) time scale~\cite{Bi_ultrafast_Fritz_Science2007, SnSe_ultrafast_Wei_npj2021,VO2_ultrafast_Li_PRX2022}. 
In Table.~\ref{table:T_elec}, we show the amplitudes of different $A_{1g}$ phonon modes found by relaxing atomic coordinates for different values of T$_{e}$. 
While the magnitude of T$_{e}$ used in these calculations is larger than the estimated maximum electron temperature in our experiment, the trend indicates that  $A_{1g}$-27 phonon mode is more likely to be excited, consistent with the experimental observations. 
Hence, in the following, we only present results for the electronic structure due to the excitation of $A_{1g}$-27 phonon mode.

Fig.~\ref{Fig4}(a) shows the calculated equilibrium band structure with all $k_z$ bands projected onto $k_{||}$ in an effort to reproduce the experimentally observed photoemission results. As expected, there is a broad feature at the top of the valence band that demonstrates the strong $k_z$ dispersion. 
There are two conduction band minima, one at the $\bar{\Gamma}$ point and the other close to the $\bar{\mathrm{Y}}$ point. The minima at the $\bar{\Gamma}$ point corresponds to $k_z=0$ whereas the one close to the $\bar{\mathrm{Y}}$ point corresponds to $k_z \approx 0.2$ (in units of $\frac{\pi}{c}$). Similarly, there are other local CBMs at $\bar{\mathrm{M}}$ and in between other high symmetry points.
%The conduction band minimum is at the $\Gamma$ point as expected, with two local minimums: one at the M point and the other in between high symmetry points.

Next, we study the evolution of band structure for different values of $Q$ corresponding to an increased population of $A_{1g}$-27 phonons. 
%Since Table~\ref{table:T_elec} shows that positive $Q$ values corresponding to A$_{1g}$-27 mode  are excited, we focus on Fig.~\ref{Fig4}(b) shows the projected band dispersion for $Q=0.6$. 
The energy cost corresponding to this value of $Q$ is $\approx$10meV/Zr atom which is of similar order of magnitude expected for a 60~$\mu J$ pump, hence we focus on only this $Q$ value here. The electronic structure for other $Q$ values are shown in the SM~\cite{Supplementary}. First, the top of the valence band moves down, while the rest of the occupied electronic structure near the $\Gamma$ point remains unchanged. This agrees with our experimental observations that the LEM shifts downwards and the TEM remains the same after photoexcitation. This shift is $\approx$20meV, agreeing with our observed shift of 10~meV. Second, the off-$\bar{\Gamma}$ conduction band minima (CBM) shift downward.
The CBM at $\bar{M}$ experiences the biggest downshift of $\approx$100 meV.
But more importantly, the CBM close to $\bar{Y}$ downshifts by $\approx$ 20 meV. This band renormalization changes the conduction band minima away from $\Gamma$ point, which would 1) prevent the electrons trapped in this band from scattering into another in conduction states and 2) grossly increase the electron-hole recombination time since $k$ is no longer preserved along the in-plane direction (increasing the overall $k$ needed for recovery).

In order to understand in more detail how the phonon-mediated change in band dispersion can affect the recombination rate of electrons and holes, we use a simple   relationship (Eq.~\ref{eq:recombination_rate}) between band edge and recombination rate proposed by Hall~\cite{hall1959recombination,varshni1967band}.
The total recombination rate is the sum of direct and indirect recombination given by the following equations:
%for the case of semiconductors. 
\begin{eqnarray}
B_d &=& \frac{A_d}{(k_BT)^{3/2}} E^2_{gd}~exp[\mathrm{min}(\frac{E_{gi}-E_{gd}}{k_BT}, 0)] ,\nonumber \\
B_i &=& A_i E_{gi} coth (\frac{\Theta}{2T}),
\label{eq:recombination_rate}
\end{eqnarray}
where, $B_d$ and $B_i$ denote direct and in-direct recombination rates and $A_d$ and $A_i$ are constants related to band mass, absorption coefficients and other parameters.  $E_{gd}$ and $E_{gi}$ are direct and in-direct band gaps. Similarly, $T$ and $\Theta$ are external and Debye temperature, respectively.
The argument in the first equation prevents exponential growth of the recombination rate when the direct band gap is smaller than the indirect band gap.

In Fig.~\ref{Fig4} (d), we plot the recombination rate for different positions of conduction band minima. The valence band maxima is fixed at the $\Gamma$ point. The first scenario shown in red curve is for the case when CBM is at the $\Gamma$ point and $E_{gi}-E_{gd} \approx 100$meV. 
This trend line shows a decrease in the recombination rates with temperature, which is expected as the increase in temperature is associated with an increased \textit{equilibrium} electron-hole pair generation. Surprisingly, the recombination rate does not change much when $E_{gi} ~\sim E_{gd}$ as shown in blue curve. This situation corresponds to our experimental condition, where we find two almost equal CBMs. 
Finally, we plot the scenario where $E_{gi} < E_{gd}$ which corresponds to the case when CBM is away from $\Gamma$ point (Fig.~\ref{Fig4} (b)). For this case, we find that the recombination rate is more than 2 orders of magnitude smaller than the first two cases. This means the photoexcited electrons at the CBM will find it harder to recombine with holes at the $\Gamma$ point compared to the case when the band gap is direct.
This supports our earlier hypothesis that phonon induced band renormalization is responsible for the loss of spectral weight of valence band at longer time delay.
In our analysis, the values of $A_d$ and $A_i$ were chosen to be 1. However in general $A_d$ is much greater than $A_i$. This would make the recombination rate even smaller for the third case.

\textit{Conclusion.--}
In summary, we performed ultrafast pump-probe ARPES measurements on a prototypical Dirac system \mater~ to reveal its electronic band structure at different time delays. 
%While electronic structure at shorter time delays were studied in the past, our studies focused on the changes in the electronic dispersion in the picosecond time delays. 
We find that the spectral weight at the top of the valence band does not recover within 10~ps suggesting 
strong electron-phonon coupling in light of the phonon-mediated metastability over 100 ps revealed in other ultrafast experiments~\cite{konstantinova2020photoinduced,HotCarrierDynamics_Li_PRB2020}.
%phonon mediated metastablity in this system consistent with other, all-optical, experimental findings that go up to much longer delay times  ~\cite{konstantinova2020photoinduced,HotCarrierDynamics_Li_PRB2020}. 
%These findings are found to be the consequence of the evolution from a direct to an indirect gap semimetal driven by light. 
The metastable states we identified in \mater~ is compatible with a quantum system operated in the frequency domain below 100 THz. Our first-principles electron and phonon calculations find that the electronic band structure is very sensitive to phonon mode induced distortions, resulting in the evolution from a direct-gap to an indirect-gap semimetal. Furthermore, our calculations suggest that even small phonon-induced band renormalization effects can reduce the electron-hole recombination rate by more than two orders of magnitude  which could provide an explanation for the missing spectral weight at the top of the valence band seen in ARPES. In our analysis, we have ignored the matrix element effects. It will be interesting to quantify the relative contributions of band renormalization and matrix element effects in future studies.

\textit{Acknowledgement.--}This work was supported by the U.S. Department of Energy (DOE) the Office of Basic Energy Sciences, Materials Sciences and Engineering Division under Contract No. DE-SC0012704.

%D. Nevola and N. Aryal contributed equally to this work.
$^\star$ Equal Contribution.
%\extrafootertext{afasf}

%\bibliography{ref}

%

\end{document}